# LIQUID COOLING OF BRIGHT LEDS FOR AUTOMOTIVE APPLICATIONS

*Yan Lai[1], Nicolás Cordero[1], Frank Barthel[2], Frank Tebbe[2], Jörg Kuhn[2], Robert Apfelbeck[2], and Dagmar Würtenberger[2]*

[1] Tyndall National Institute, Cork, Ireland
[2] Schefenacker Vision Systems GmbH, Schwaikheim, Germany

## ABSTRACT

With the advances in the technology of materials based on GaN, high brightness white light emitting diodes (LEDs) have flourished over the past few years and have shown to be very promising in many new illumination applications such as outdoor illumination, task and decorative lighting as well as aircraft and automobile illuminations. The objective of this paper is to investigate an active liquid cooling solution of such LEDs in an application of automotive headlights. The thermal design from device to board to system level has been carried out in this research. Air cooling and passive liquid cooling methods are investigated and excluded as unsuitable, and therefore an active liquid cooling solution is selected. Several configurations of the active liquid cooling system are studied and optimisation work has been carried out to find an optimum thermal performance.

## 1. INTRODUCTION

Due to the small package size, styling flexibility and superior performance over incandescent light sources, LEDs are widely used in many automobile exteriors nowadays, such as brake lights, turn indicators and tail lights. With the development of higher light output packages, the use of white LED sources for vehicle forward lighting applications is beginning to be considered. Although many properties of LEDs have made them a very promising light source for vehicle forward lighting, the use of white LEDs as automobile headlamps is in its infancy. Currently, LEDs have appeared as forward lighting only in some concept cars, and there are no LEDs customised for headlight applications.

At present, LEDs offer a high cost solution with insufficient lumen output for production vehicles. Legal requirements stipulate that 750 lm per lamp is required for headlamps. However, since the average current bright LED output is only 40 lm/W, more LEDs and higher driving powers are needed to meet this standard [1].

As the demands for light output increases, the driving power of the LED increases continuously. The thermal management of LED packaging, which has great effect on their efficiency, performance and reliability, has become more and more important for these devices.

As a result of an increase in diode junction temperature there is a decrease in the LED efficiency and a shift in the emission wavelength. Therefore, the LED operating temperature must be kept well below its maximum operating temperature (e.g. < 125 °C) for optimum efficiency operation and small colour variation. To achieve this, the thermal solution must be all-inclusive and must address thermal issues at all levels—device, package, board and system level. Commercially available bare die bright LEDs are used in this application. Thermal simulations using Computational Fluid Dynamics (CFD) were carried out at all levels to support the search for a suitable thermal management solution. The design of the thermal management solution was supported using the commercial CFD software FloTherm [2].

## 2. CHOICE OF ACTIVE LIQUID COOLING

### 2.1. From device to board level

The chosen LED for this application is a Cree XBright900. This LED is a 900×900 µm chip that is commercially available as bare die [3]. It generates light of wavelengths between 460 and 470 nm in 2.5 nm range bins, which gives the colour of blue. With apropriate thermal management it can operate generating up to 2.7 W of heat per LED. The system proposed here consists of 15 LEDs mounted on 5 boards with 3 LEDs each.

To simplify the mounting process, the LEDs had to be individually packaged. Furthermore, the LEDs need to incorporate a layer of phosphor to convert the blue light from the GaN based LED to a white (visible spectrum) light emitter. The heat is dissipated directly from the active region of the device to the package. Therefore, a high thermal conductivity ceramic must be chosen to provide the package with a low resistance thermal path and electrical insulation. AlN (k=200 W/mK), in this case, fits the role very well in providing conduction and heat spreading for high power operation. The calculated thermal resistance between the LED and the bottom of the AlN package is less than 2 °C/W [4].

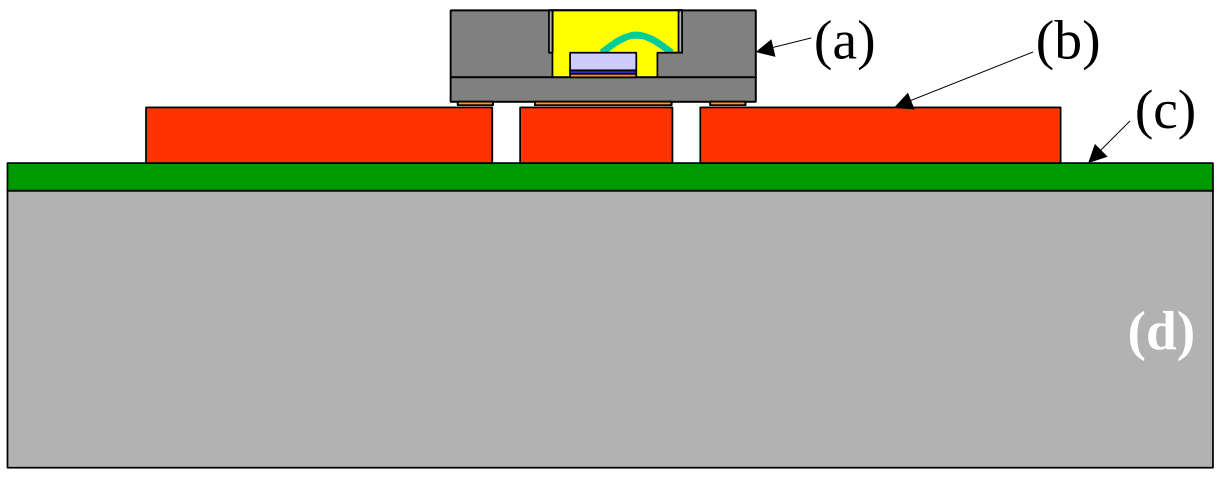

***Figure 1. Insulated Metal Substrate assembly. (a) AlN cup with wire-bonded LED, (b) Circuit layer, (c) Dielectric layer and (d) Aluminum substrate.***

The AlN package is then mounted on an insulated metal substrate (IMS) (*Figure 1*). Adopting IMS provides both heat spreading and a good thermal path to the heat sink and greatly simplifies the system design. IMS consists of three layers: a copper foil circuitry layer bonded together with a thin dielectric layer, and a metal base plate made of aluminium [5].

Several different materials making up the dielectric layer as well as different combinations of the thickness of the three layers of IMS have been compared. The thermal simulations show that the optimum board should have a thick circuitry layer to spread the heat while a very thin dielectric layer made of a material with higher thermal conductivity to reduce the thermal resistance. The thicknesses of these layers are therefore only limited by the manufacturability of the IMS. The chosen structure of IMS includes a 70 µm copper layer, a 75 µm dielectric layer with thermal conductivity of 2.2 W/mK and a 1 mm Al core board (*Table 1*) [4].

| Layer | Material | Thickness | k (W/mK) |
|---|---|---|---|
| Circuit | Cu | 70 µm | 385 |
| Dielectric | Ceramic/Polymer | 75 µm | 2.2 |
| Metal Substrate | Al | 1 mm | 200 |

***Table 1. IMS board structure and materials used in modelling.***

### 2.2. System level -- air cooling

The headlight application requires forward light emission. To achieve this, the optical design requires the IMS boards to be mounted at 45° facing the back of the headlight assembly. For passive air-cooling, the heat sink is mounted directly behind the IMS board. In the real application, the whole system is placed inside the headlamp enclosure, which reduces the heat dissipation by convection. Due to the space limitation inside the headlamp, the size of the heat sink is restricted. As a result (*Figure 2*), the LED junction temperature far exceeds its maximum value of 125°*C*.

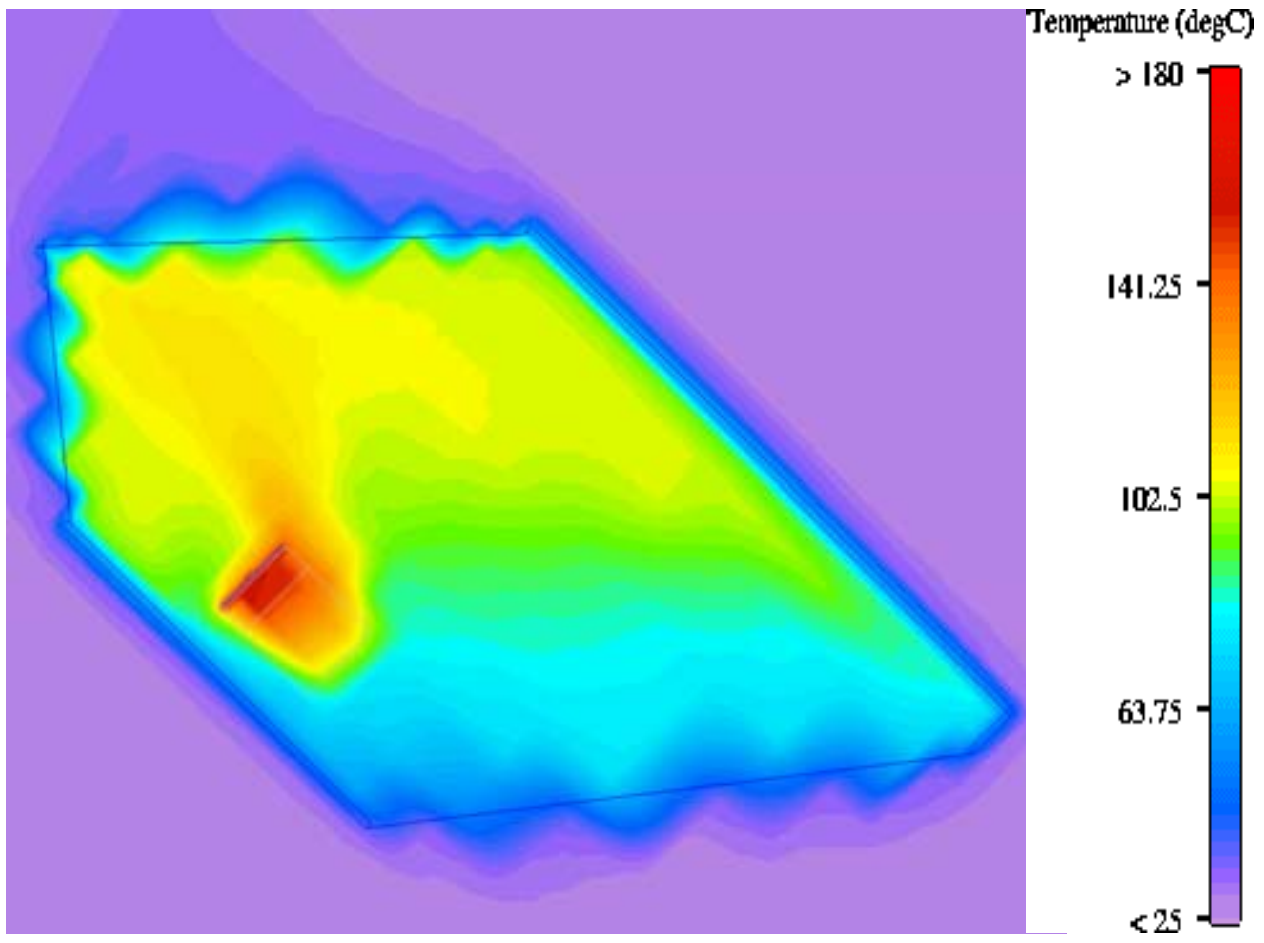

***Figure 2. Temperature profile across the headlight assembly for passive air cooling ($T_j$=200°C).***

Active air cooling is also investigated here. However, it is not a feasible cooling solution since the space and enclosure constraints would necessitate a large number of high flow fans. This is impractical from reliability, cost and assembly viewpoints. Therefore, liquid cooling solutions are chosen for further analysis.

### 2.3. System level – Passive liquid cooling

Two possible passive liquid cooling configurations have been investigated: passive closed-loop and heat pipe.

Simulations show that a passive closed-loop could achieve the required cooling levels to keep the LED junction temperatures well below their maximum operation temperature. However, in passive systems the motion of the fluid is achieved by buoyancy forces. Therefore, these systems require the heat exchanger to be placed above the heat source, in which fashion the hotter and lighter fluid (water) will travel upwards against gravity to be cooled. However, although feasible from the thermal point of view, it is not a suitable solution for the cooling of headlamps, as the headlight design requires the heat exchanger to be positioned below the LED modules.

For the heat pipe solution, a loop heat pipe system is applied in order to circulate the system However, since in this application each LED board needs to be adjustable individually, the heat pipe is therefore required to be flexible, which significantly increases the cost of the cooling solution. From the available flexible heat pipe products (e.g. Thermotek, Dau), the price could be as expensive as $1,000 per unit. Again, although feasible from the thermal point of view, the loop heat pipe system

is not a suitable solution due to engineering and cost considerations.

Therefore, the cooling solution of high brightness LEDs in the automotive application turns to active liquid cooling.

## 3. ACTIVE LIQUID COOLING

### 3.1. System structure

The liquid cooling system consists of a pump, cold plates thermally connected to the heat sources (IMS boards), a reservoir, and a heat exchanger. The components are connected by flexible hoses which create a closed loop.

As each board has to be individually adjustable, a separate cold plate is attached to each board. Due to the weight and volume limitations, and the fact that the high and low beam are never on at the same time, a single heat exchanger can be shared by both the high beam (HB) and low beam (LB). In this way, the size of the heat exchanger can be doubled and therefore more heat dissipation is provided. The heat exchanger consists of a heat sink with a liquid-cooled base. Due to its good thermal properties and availability, the most suitable liquid for the cooling solution is water with a number of additives (e.g. antifreeze -glycol-, anti-algae, anti-fungal, etc).

Several configurations are considered here. To reduce damage to the pump and therefore improve its reliability, the pump only sees ‘cold’ liquid. First a solution with five LB-HB circuits in parallel (*Figure 3*) is investigated. Although optimal from the thermal point of view, this solution requires two manifolds plus two different hose sections, which makes the system far too complicated and therefore, not the best option in this case.

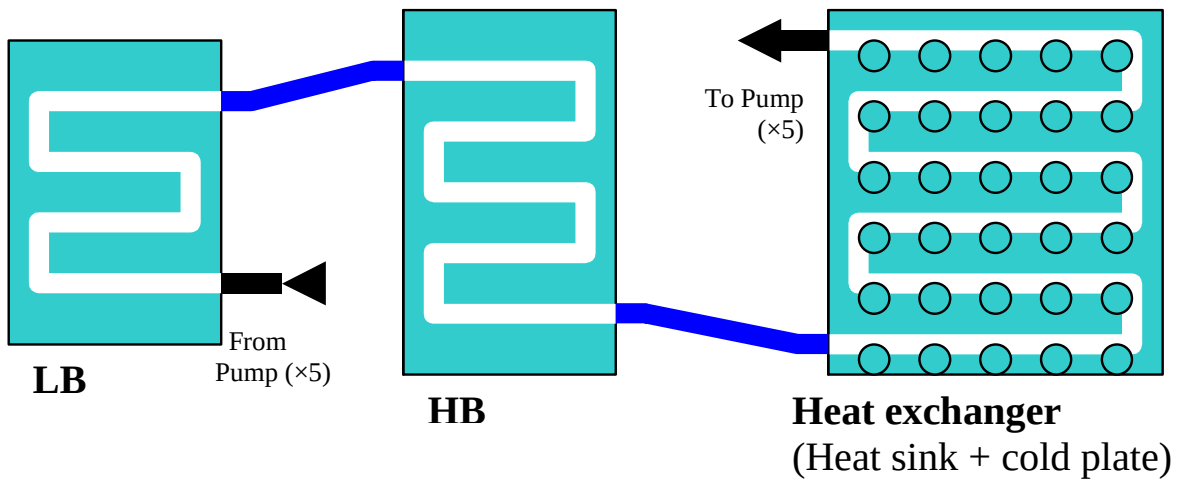

***Figure 3. Cold plates and heat exchanger design and their hose connections.***

A second solution consists of the same five LB-HB cold plate circuits but connected in series on a single loop. The circuit is longer and therefore the pressure drop is higher. Thermal simulations show that the pressure drop in the circuit is well below the pressure head of standard pumps and therefore should not have a detrimental effect on the thermal performance of the liquid cooling solution.

Finally, an alternative design is proposed, consisting of a liquid loop through all the LB cold plates in series followed by the HB in series and then into the heat exchanger (*Figure 4*). This design presents the advantages of having a smaller number of hoses (14 instead of 17), shorter hoses which allows the separate adjustment of both beams, and being simpler to mount. The thermal simulations show that the junction temperature of the last set of three LEDs (last board in the loop) is less than 5°C higher than those in the first set.

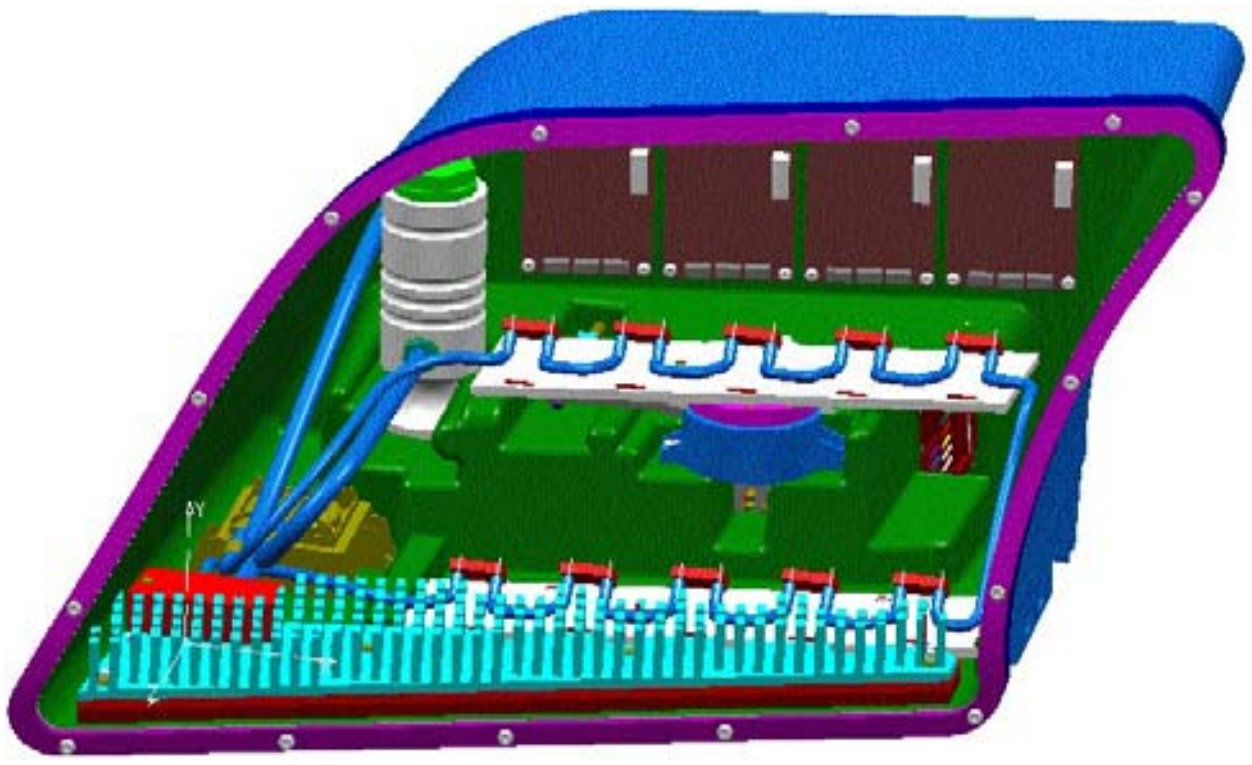

***Figure 4. Active liquid cooling configuration—the liquid loop connects all the LB cold plates in series followed by the HB in series and then into the heat exchanger.***

The system configuration modelled using Flotherm is shown in Figure 5.

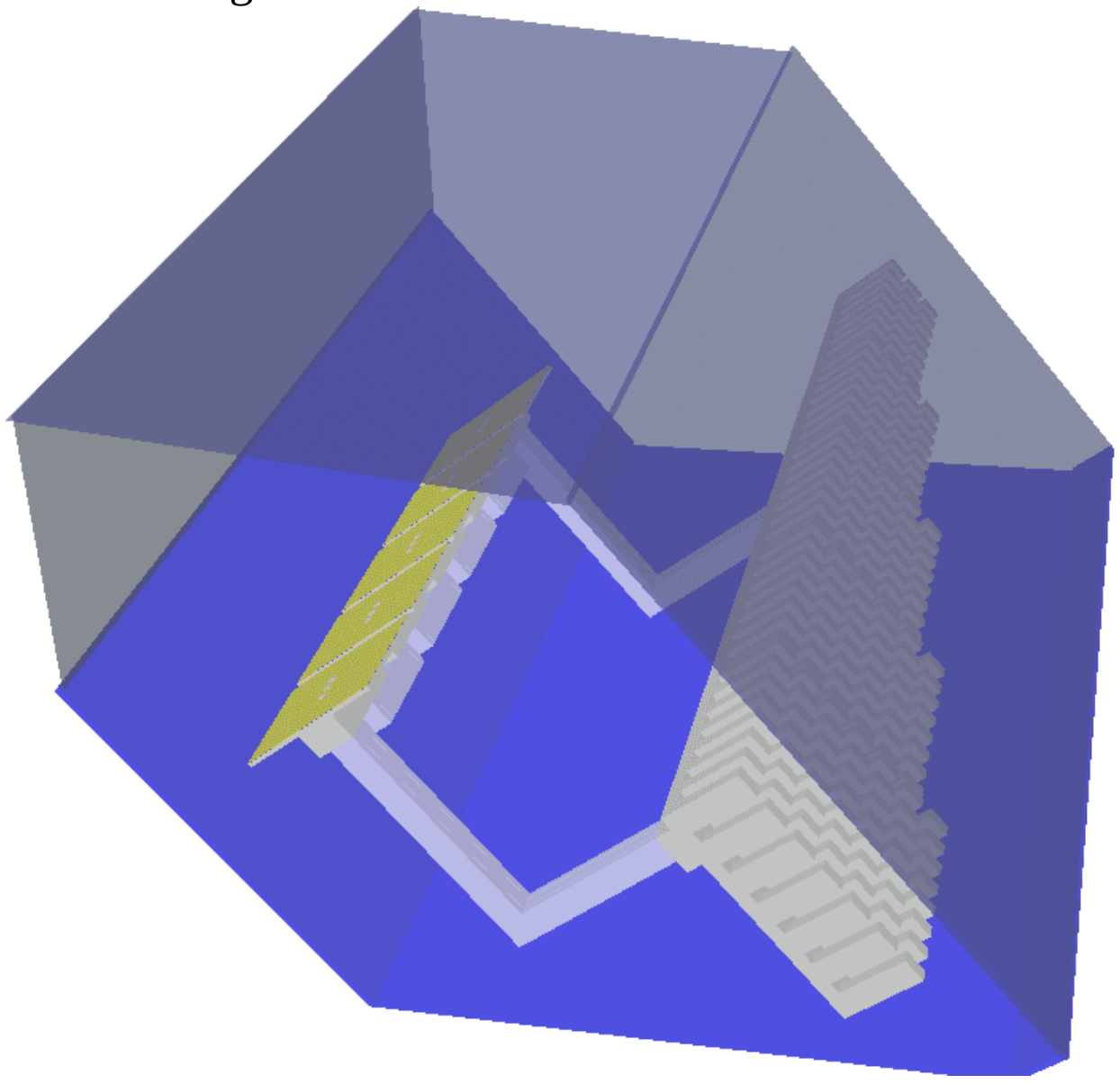

***Figure 5. Full model of active liquid cooling of complete low beam system inside the headlamp enclosure (shown in Figure 4).***

### 3.2. Thermal optimisation

*3.2.1. Liquid flow optimisation*

*Figure 6* shows the calculated LED temperature as a function of the nominal (zero pressure) pump flow. As the nominal flow of the pump increases, the LED junction temperature decreases. However, for flows higher than 0.12 l/s, the reduction is less significant.

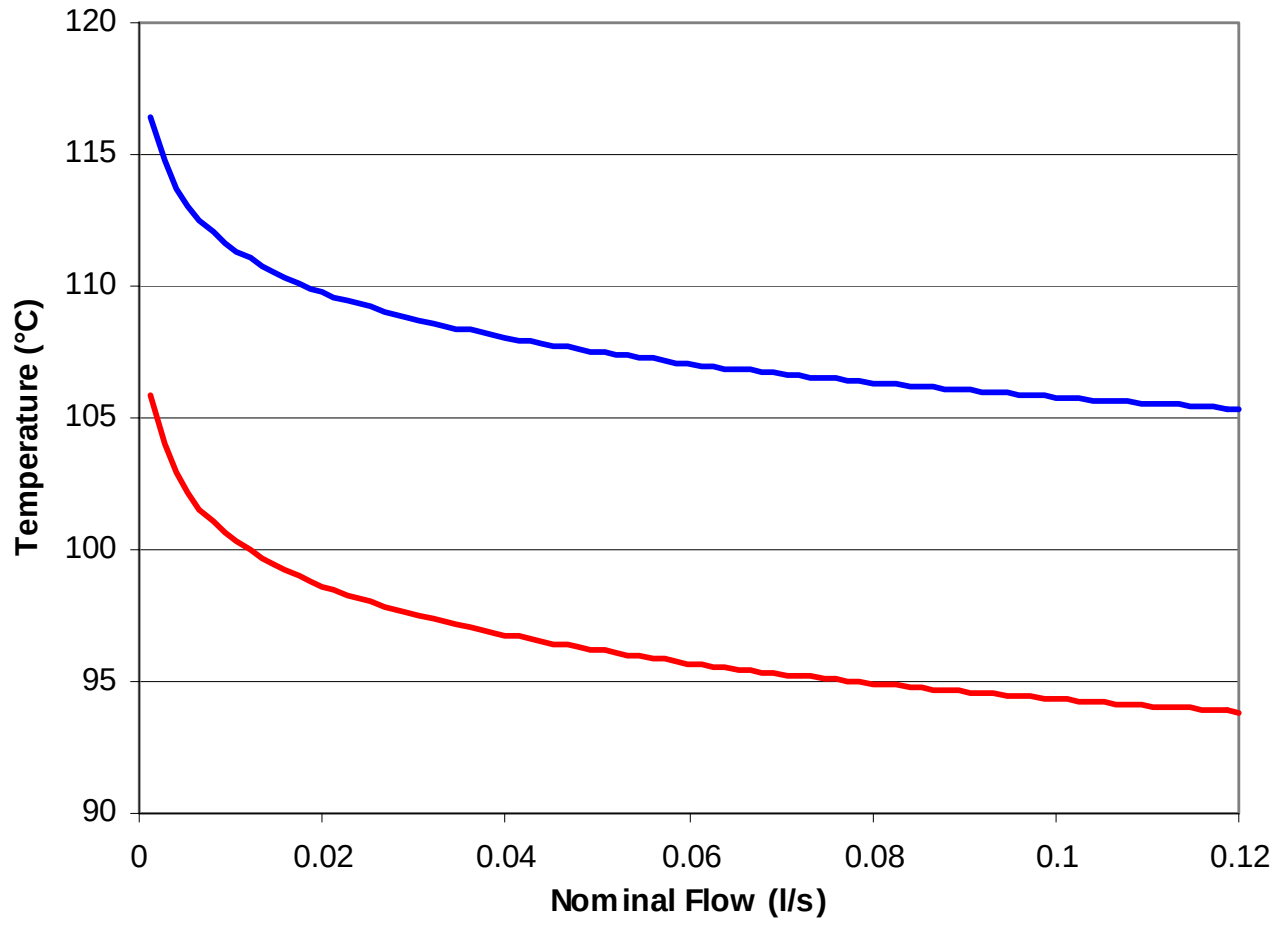

**Figure 6. *Calculated LED junction temperature (blue) and IMS board temperature (red) as a function of the nominal flow (with a nominal pressure head of 25 kPa)***

*Figure 7* shows the calculated relation between the nominal and the actual flows. For low nominal flows the effect of the pressure drop is insignificant. However, as the flow increases, the pressure drop in the liquid cooling loop limits the actual flow. *Figure 8* shows the relation between pressure drop and flow for the loop and the characteristics of a linear pump with a nominal flow of 0.12 l/s and a nominal pressure head (no flow) of 25 kPa. The results show that the chosen pump will be operating within its recommended operation range.

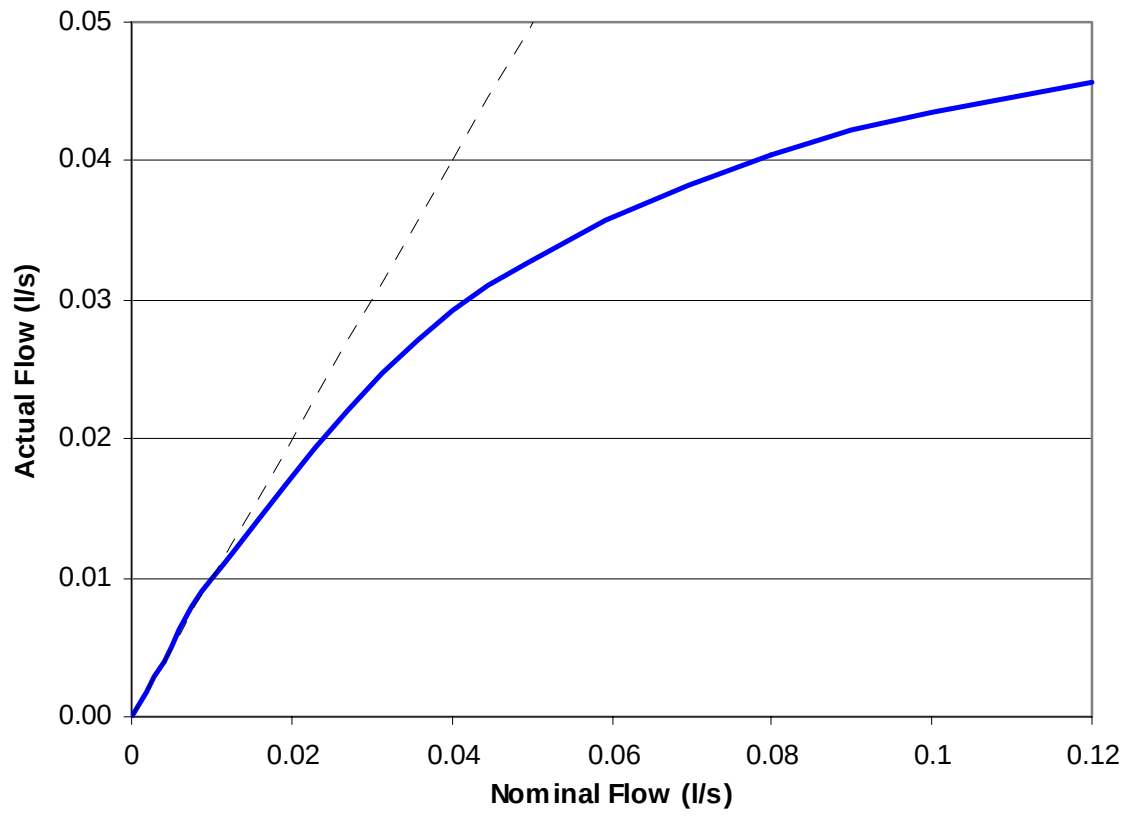

**Figure 7. *Calculated actual flow as a function of the nominal (zero pressure) flow of the pump.***

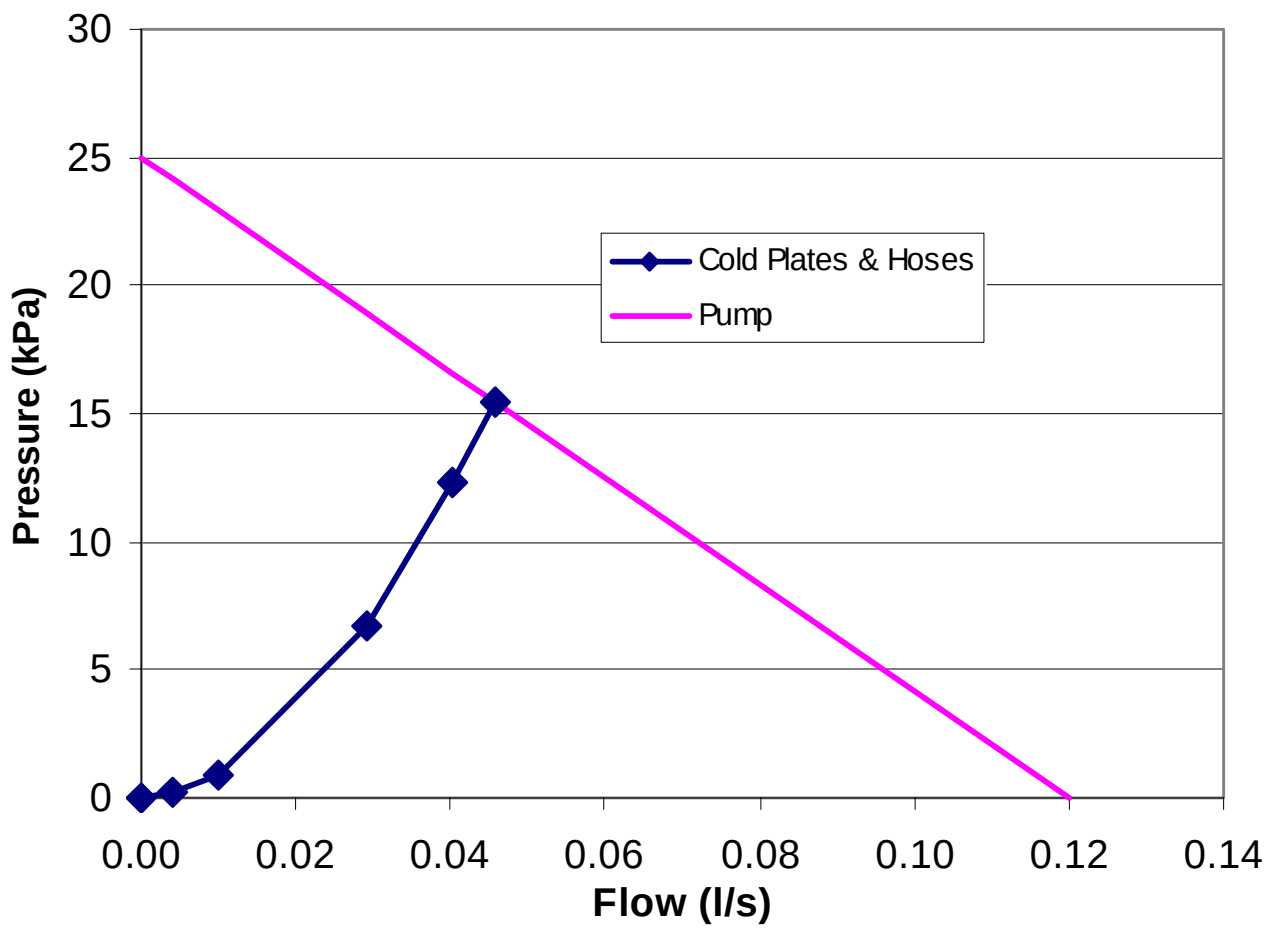

**Figure 8. *Pressure vs. flow characteristics for the liquid cooling loop and linear pump characteristics***

*3.2.1. Heat exchanger (heat sink) optimisation*

The heat sink design relies on the external condition of the heat sink, such as the airflow type and the operating environment that determines the components placement and the air flow rate [6]. In this application, since the heat sink is placed horizontally, there is no preferential flow direction, and therefore, in order to reduce the total weight, the pin configuration is selected for the heat sink.

There are a number of parameters to be considered in the design of the optimal heat sink, such as pin length, pin numbers, base thickness etc. Due to the conflicting effects of some of the parameters on the LED temperature, investigation of these parameters is undertaken through an iterative procedure.

The parameters studied in this case include the following (see Figure 9):

1) Heat sink base thickness (***t***). Since the cold plate underneath has already spread the heat over the whole area, the base thickness has very small influence on the LED temperature. Due to the weight constraints, it should be as thin as mechanically possible. For the rest of the optimisation, it is set to 5 mm.
2) Heat sink height (***H***). The total height of the heat sink is equal to the base thickness (t) plus the pin height (h). The pin height is the most dominant parameter in the optimisation. Therefore, from the thermal point of view, it should be as high as possible within the limitation that the heat sink should not block the out-coming light.
3) Pin length (***l***). The calculated optimum value is 4.5 mm. However, the sensitivity of the LED temperature on pin length is small for values around this optimum. According to the simulations, the values between 3.5 and 6 mm only contribute to an increase of less than 1 °C. Therefore, any value in this range is suitable in this application.
4) Pin width (***w***). The calculated optimum value is 9mm. As the same with the case of pin length, small variations in the range between 7.5 and 10 mm only produce a small increase in temperature.
5) Number of pins in the X direction ($N_x$). The calculated optimum value is 40, which corresponds to spacing between pin of 5.1 mm. Again, from the thermal simulations, any number falls in the range between 35 and 45 works in this case.
6) Number of pins in the Y direction ($N_y$). Due to the limited space in the headlamp, the maximum number referred to here is on the widest edge, the number will be smaller in the narrow end. The calculated optimum value is 7, which corresponds to a spacing of 4 mm. Values from 7 up are also feasible.
7) The total weight of this optimum heat sink realised in aluminium is less than 800 grams.

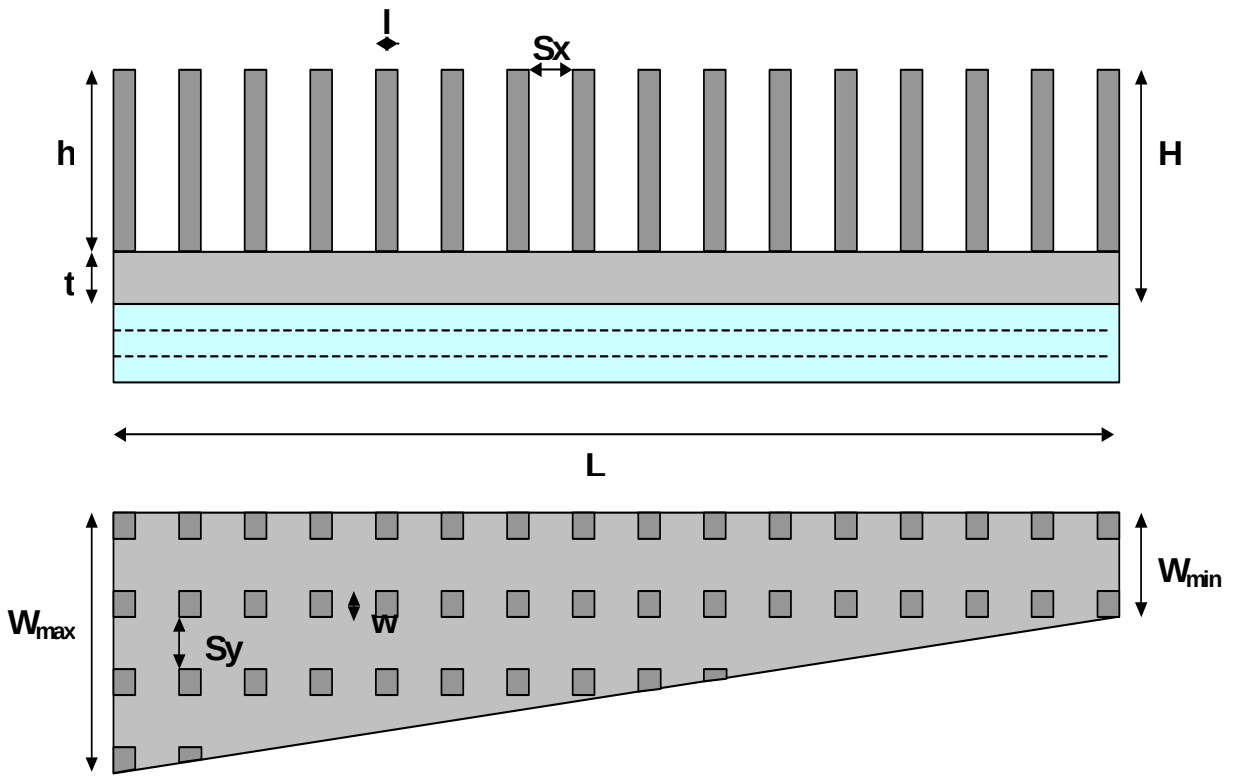

***Figure 9. Heat sink parameters and dimensions***

The optimisation of some parameters depends on others (e.g. pin length and pin width). Therefore, they had to be considered simultaneously during the optimisation process (see Figure 10).

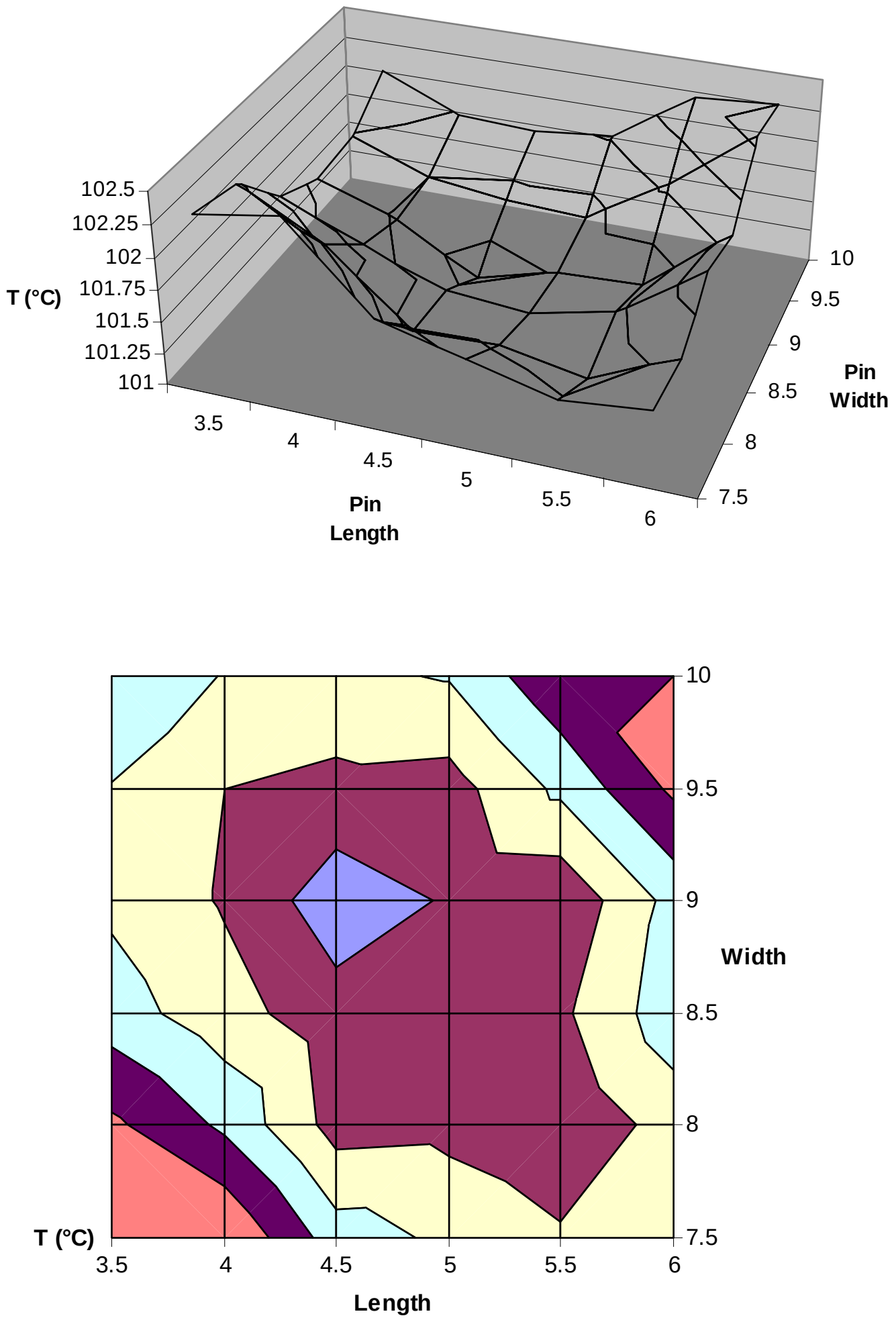

***Figure 10. LED temperature as a function of pin length and pin width. A) 3D view and B) Contour plot.***

Other parameters (e.g. pin number) are independent and can therefore be optimised separately (Figure 11).

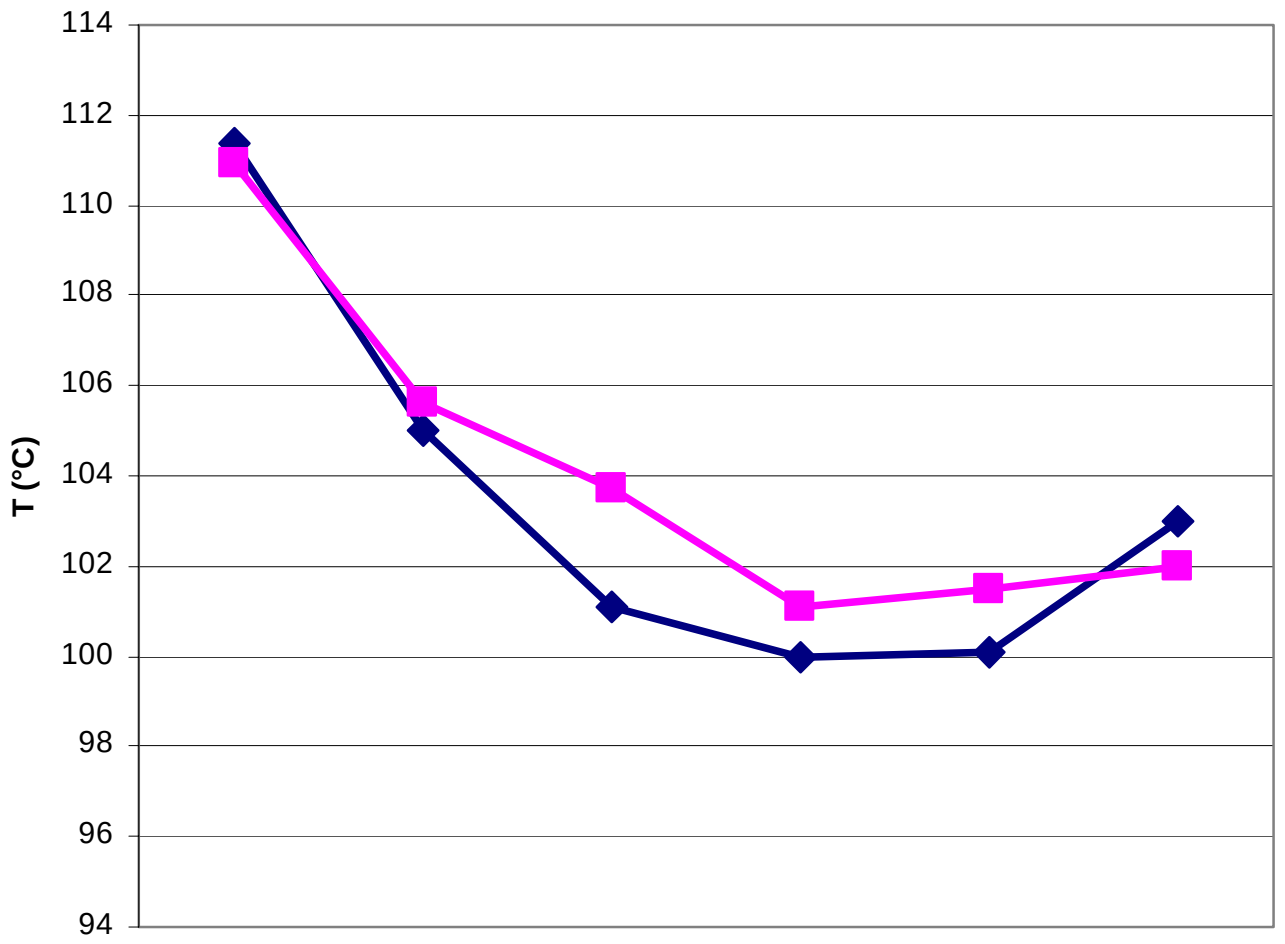

***Figure 11. LED temperature as the function of pin number in the X (diamonds) and Y (squares) directions.***

In summary, the dimensions of the optimised heat sink for this application are as follows (all dimensions in mm).

- Optimised design
  t = 5, H > 30, h > 25, l = 4.5, w = 9, $N_x$=8, $N_y$=7
- Allowable design margins ( < 1 °C increase in junction T)
  3.5<l<6, 5<w<10, 35<$N_x$<45, 7<$N_y$

## 4. CONCLUSIONS

This paper demonstrates the procedure for selection and optimisation of an active liquid cooling solution for high brightness LEDs customised for novel headlight applications.

It was found that air and passive liquid cooling was either insufficient to maintain LED junction temperature below its maximum allowable levels or unfeasible to realise in the actual application. While some of these solutions would be suitable from a purely thermal point of view, this is not the case when the optical and mechanical designs are taken into account. Therefore all aspects of the headlight design must be taken into account when seeking a suitable thermal management solution.

Therefore active liquid cooling is selected as the optimum cooling solution under these circumstances. Several different system structures of active liquid cooling are studied and compared in this paper. And thermal optimisations of the liquid flow and heat sink are carried out in order to maximise the thermal performance. During the search for the optimum thermal solution, thermal management is not the only factor to focus on; all related issues such as manufacturability and product specifications are also taken into account.

With the development of brighter white LEDs, the driving power required for a certain light output will be decreasing continuously in the future. Therefore heat dissipation will also decrease. With the reduced power requirements for the system and lower heat dissipation, the cooling solution can once again be simplified to only passive air-cooling.

## 5. ACKNOWLEDGEMENTS

This work was partly supported by the European Commission through the FP6-Transport project ISLE (Contract TST3-CT03-506316). The authors would like to thank Rafael Jordan (TU Berlin) and Jochen Kunze (Global Light Industries GmbH) for fruitful discussions and their support of this work